\documentclass[english,aps,preprint,nofootinbib]{revtex4}%
\usepackage[T1]{fontenc}
\usepackage[latin9]{inputenc}
\usepackage{amsmath}
\usepackage{amssymb}
\usepackage{babel}
\usepackage{amsfonts}
\usepackage{graphicx}
\usepackage{physics}

\begin{document}
\title{Wigner Friend scenarios with non-invasive weak measurements}
\author{A. Matzkin}
\affiliation{Laboratoire de Physique Th\'eorique et Mod\'elisation, CNRS Unit\'e 8089, CY
Cergy Paris Universit\'e, 95302 Cergy-Pontoise, France}
\author{D.\ Sokolovski}
\affiliation{Departmento de Qu\'imica-F\'isica, Universidad del Pa\' is Vasco, UPV/EHU,
Leioa, Spain}
\affiliation{IKERBASQUE, Basque Foundation for Science, E-48011 Bilbao, Spain}

\begin{abstract}
Wigner Friend scenarios -- in which an external agent describes quantum
mechanically a laboratory in which a Friend is making a measurement -- give
rise to possible inconsistencies due to the ambiguous character of quantum
measurements. In this work, we investigate Wigner Friend scenarios in which
the external agents can probe in a non-invasive manner the dynamics inside the
laboratories. We examine probes that can be very weakly coupled to the systems
measured by the Friends, or to the pointers or environments inside the
laboratories. These couplings, known as Weak Measurements, are asymptotically
small and do not change the outcomes obtained by the Friends nor their
probabilities. Within our scheme, we show that the weakly coupled probes
indicate to the external agents how to obtain consistent predictions,
irrespective of the possible inconsistencies of quantum measurement theory.
These non-invasive couplings could be implemented with present-day technologies.

\end{abstract}
\maketitle

\section{Introduction}

Measurements are special \ in Quantum Mechanics. Indeed quantum systems evolve
unitarily, except when a measurement takes place: a \textquotedblleft special
rule\textquotedblright\ needs to be enforced in order to account for the
appearance of a definite measurement outcome out of the linear superposition
of pointer-system states. Technically, this rule appears as a projection of
the quantum state right before measurement to an eigenstate of the measured
observable. Physically, there is no unambiguous and consistent manner of
accounting for this rule \cite{wheeler-zurek}.\ Logically, this makes quantum
theory inconsistent \cite{deutsch}. Several options are on the table, but none
of them is entirely satisfactory and there is unsurpisingly no consensus on
these issues, overall known as the celebrated Measurement problem.

Could there be an underlying basis that would account for the projection that
takes place when a measurement is completed -- the ``collapse'' of the
wavefunction ? In a series of papers \cite{wigner-r,wigner-margenau,wigner69},
Wigner assumed unitary evolution to be universal -- implying it also applies
to macroscopic bodies -- but that linear superposition cannot be applied to
conscious observers making a measurement. Hence if a Stern-Gerlach experiment
takes place in a sealed laboratory L, an agent W outside the laboratory would
describe the state of the atom having passed through the inhomogeneous
magnetic field by a linear superposition of up and down spin along the field
direction. Now if there is a friend F inside the sealed lab running the
experiment, she would take the superposition to \textquotedblleft
collapse\textquotedblright\ upon completion of her measurement. But how should
W describe the experiment?

If unitary evolution is universal, the answer would be that as long as W is
unaware of the outcome, he will take the quantum state of the laboratory to be
in superposition even after F completed her measurement. Wigner was dubious
that a superposition of different states of consciousness would make
sense\footnote{Actually Wigner's position was that quantum mechanics does not
apply to conscious Observers. He speculated \cite{wigner-r,wigner69} that
consciousness induces a non-linear modification of linear quantum mechanics,
though he will change his mind afterwards \cite{wigner-late}.}, but concluded
(at least in his 1962 paper \cite{wigner-r})) that an experimental test
attempting to distinguish a linear superposition of L from a mixed state would
in practice be impossible. Still it is worthwhile to investigate whether such
assumptions can be taken to be consistent.

Remarkably, in two recent works Brukner \cite{bruckner} and Frauchiger and
Renner \cite{renner} have shown by employing an extended Wigner friend
scenario composed of two sets $i=1,2$ each involving a Friend F$_{i}$ inside a
Laboratory L$_{i}$ and an agent W$_{i}$ external to the laboratories L$_{1} $
and L$_{2}$ that assuming the Friends get a definite outcome while the agents
W$_{i}$ describe these processes by unitary evolutions leads to a
contradiction. The implications of this has sparked a wild interest
\cite{sudbery,laloe,drezet,dustin,lombardi,wiseman,relano,zukowski}, though
its relevance depends on a set of specific assumptions that are explicitly or
implicitly being made.\ 

In this work we will not be interested in these specific assumptions -- in
particular in those made in \cite{bruckner,renner} -- but focus instead on the
basic hypothesis at the root of the Wigner Friend Scenarios according to which
a completed measurement in a closed laboratory can be described by an external
agent as a unitary evolution. We will introduce a procedure by which the
agents W$_{i}$ can acquire information on the state of the system or the state
of the pointer inside the laboratories L$_{i}$ without perturbing in any
essential way the measurements taking place in L$_{i},$ in particular without
modifying any outcome probabilities. This information will be acquired with
non-invasive probes through a minimal disturbing scheme known as weak
measurements \cite{AAV}. In general, a weak measurement involves a unitary
interaction between a system and a very weakly coupled probe. The weakly
coupled probe becomes entangled with the system, and it is only when the
system is subsequently subjected to a measurement of another observable that
the information can be retrieved from the probe.

Here our strategy will be to connect a non-invasive probe available to the
agents W$_{i}$ to elements inside the otherwise sealed laboratories L$_{i}%
$.\ For instance such probes may be coupled through a weak interaction to the
spins inside L$_{i},$ to the pointers measuring the spins, or event to the
environment inside L$_{i}$ by which the Friends acquire information on the
spin measurement outcomes. We will see that even in the original Wigner setup,
assuming unitary evolution leads to a contradiction as W would read from the
weakly coupled probe that the spin (or the pointer that measured the spin, or
the environment inside L) is in a state of superposition, while F announces
she obtained a definite outcome. This contradiction subsists in the extended
scenario: the Friends announce they obtained a definite outcome but if the
external agents assume unitary evolution, this would not be reflected in the
state of the probes.\ Actually if the agents W$_{i}$ are allowed to
communicate, they would realize their probe states act as a witness of
entanglement between the spins (or the pointers, or the environments) of each
laboratory. This contradiction is of course the counterpart of the
contradiction uncovered in Refs. \cite{bruckner,renner}. As expected, no
contradiction occurs if the Friends' measurements result in a global collapse:
in this case the weakly coupled probes are in perfect agreement with the
outcomes announced by F$_{1}$ and F$_{2}$. Note that in this paper we will
take \textquotedblleft collapse\textquotedblright\ to represent the second
step of the measurement process, recalled in Sec. \ref{sec2}, without any
commitment to an underlying picture nor specific interpretation (we will
sometimes recall this point by writing \textquotedblleft possibly effective
collapse\textquotedblright).

In Sec.\ \ref{sec2} we will introduce the original (WFS) and the extended
Wigner Friend (EWFS) setups we will be working with. Sec. \ref{sec3} will deal
with implementing\ different non-invasive measurements with weakly coupled
probes in the laboratories for both the WFS and EWFS. We will assume the
external agents W can compute the quantum state evolution either by following
full unitary evolution or by applying globally the projection postulate, each
case leading to different predictions for the weakly coupled probes. We will
discuss our results in Sec. \ref{discussion}; in particular we will argue that
although connecting a weak non invasive probe to a laboratory breaks the
independence of L from a logical viewpoint, it does not change the physical
situation in the laboratories. We close with our conclusions in Sec.
\ref{sec5}.

\section{Wigner Friend setups}

\label{sec2}

\subsection{WFS: Single laboratory and observers \label{1L}}

\begin{figure}[tb]
\includegraphics[width=8cm]{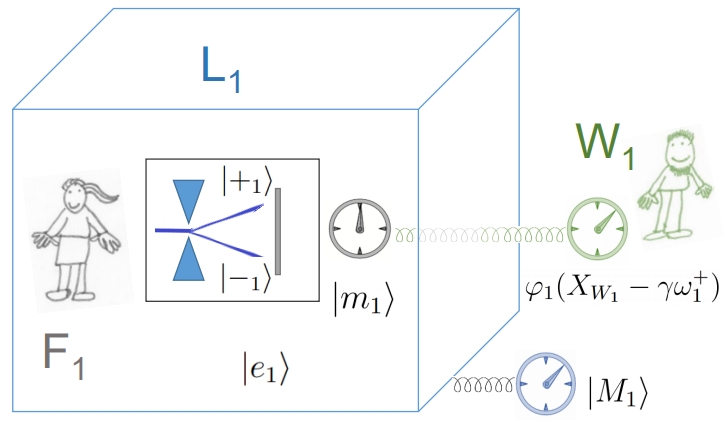}  \caption{Wigner Friend Scenario (WFS).
In the original WFS, the Friend F$_{1}$ measures the spin of an atom with her
pointer in state $\ket{m_1}$ inside the sealed laboratory L$_{1}$. The
external agent W$_{1}$ measures L$_{1}$ with his pointer in state $\ket{M_1}$.
In this work, W$_{1}$ has in addition a non-invasive probe [online color: in
green], coupled here to the Friend's pointer. After F$_{1}$ completes her
measurement, the probe, initially in the quantum state $\varphi_{1}(X_{W_{1}%
})$, will have shifted to $\varphi_{1}(X_{W_{1}}-\gamma\omega_{1}^{-})$ where
$\omega_{1}^{-}$ is the so-called weak value.}%
\label{fig1}%
\end{figure}

In the original Wigner Friend scenario (WFS) Wigner \cite{wigner-r} introduces
a laboratory L in which a friend F performs a Stern-Gerlach experiment on an
atomic spin, while an agent W is outside the isolated laboratory and
ultimately measures the quantum state of the laboratory (see Fig. \ref{fig1}).
We will introduce the following quantum states:
\begin{align}
\left\vert \Psi(t=0)\right\rangle  &  =\underbrace{\left\vert
s(t=0)\right\rangle \left\vert m(t=0)\right\rangle \left\vert
e(t=0)\right\rangle }\left\vert M(t=0)\right\rangle \label{w1}\\
&  \equiv\qquad\qquad\left\vert L(t=0)\right\rangle \qquad\otimes
\qquad\left\vert M(t=0)\right\rangle . \label{w2}%
\end{align}
$\left\vert s\right\rangle $, $\left\vert m\right\rangle $ and $\left\vert
e\right\rangle $ label respectively the spin, the pointer measuring the spin,
and the environment of the isolated laboratory, whose overall quantum state is
denoted by $\left\vert L\right\rangle $. $\left\vert M\right\rangle $ is the
quantum state of W's pointer, that is assumed to measure L.

A measurement in quantum theory is represented in two steps. First a unitary
evolution
\begin{equation}
\exp\left(  -i\int_{0}^{t}dt^{\prime}H_{int}(t^{\prime})/\hbar\right)
\label{uc1}%
\end{equation}
couples the system observable $\hat{A}$ that is being measured to a pointer
observable $\hat{P}$ (typically the momentum). The coupling Hamiltonian is
$H_{int}=g(t)\hat{A}\hat{P}$ where $g(t)$ is the coupling strength. Before the
measurement the overall quantum state can be put as $\psi(t=0)=\sum_{k}%
c_{k}\left\vert a_{k}\right\rangle \left\vert m\right\rangle $, where the
system is expanded over the eigenstates $\left\vert a_{k}\right\rangle $ of
$\hat{A} $. The pointer state $\left\vert m\right\rangle $ is typically a
tightly localized function (eg, a Gaussian), initially centered on $X=0$.
Unitary evolution leads to
\begin{align}
\psi(t)  &  =\exp\left(  -ig\hat{A}\hat{P}/\hbar\right)  \psi(t=0)
\label{vN0}\\
&  =\sum_{k}c_{k}\left\vert a_{k}\right\rangle m(X-ga_{k}) \label{VNe}%
\end{align}
where we have $\left\langle X\right\vert \left.  m_{k}\right\rangle
=m(X-ga_{k})$ and set $g\equiv\int dt^{\prime}g(t^{\prime})$ assuming $\hat
{A}$ is time independent and recalling that $\exp\left(  -iga_{k}\hat{P}%
/\hbar\right)  $ is a translation operator. Each term of the sum in Eq.
(\ref{VNe}) represents the eigenstate $\left\vert a_{k}\right\rangle $
correlated with a pointer state shifted by $ga_{k}$ relative to the original
position. For sufficiently strong couplings $g$ the pointer wavefunctions
$m(X-ga_{k})$ do not overlap in position space so measuring the pointer
position directly yields the corresponding eigenvalue.
%ipso-facto ``collapses'' the entangled superposition (\ref{VNe}) to a
%single summand.

In the Wigner Friend scenario, F measures the spin component along some
direction, say $\mathbf{x}$, hence $\hat{A}$ is given by $\sigma_{x}$ and
there are 2 pointer states $m(X\mp g)$ correlated resp. with the spin
$s_{x}=\pm1$. The second step of the measurement is the projection that, as
was first axiomatized by von Neumann \cite{neumann}, is necessary in order to
account for the appearance of a definite outcome. Indeed including additional
unitary interactions, eg between the pointer and the environment of the
pointer (that could include the photons having interacted with the pointer
traveling toward the Friend's eye) would only create a growing chain of
entangled states, the von Neumann chain, defined in the measurement basis; eg,
including the quantum states of the environment $\left\vert e\right\rangle $
is tantamount to replacing Eq. (\ref{VNe}) by $\sum_{k}c_{k}\left\vert
a_{k}\right\rangle \left\vert m_{k}\right\rangle \left\vert e_{k}\right\rangle
$.

So as far as F measuring the spin\ is concerned, the inital state $\left\vert
L(t=0)\right\rangle $ introduced in Eq. (\ref{w2}) goes into $\left\vert
L(t)\right\rangle =\sum_{\pm}c_{\pm}\left\vert \pm\right\rangle m_{\pm
}\left\vert e_{\pm}\right\rangle $ in the first unitary step but once the
measurement is completed at $t=t_{f}$ and an outcome, say $+$ (resp.\ $-$) is
obtained, the final state is the projection $\left\vert \mathcal{L}%
+\right\rangle \equiv\left\vert +\right\rangle \left\langle +\right\vert
\left.  L(t_{f})\right\rangle $ (resp. $\left\vert \mathcal{L}-\right\rangle
\equiv\left\vert -\right\rangle \left\langle -\right\vert \left.
L(t_{f})\right\rangle $). According to the \textquotedblleft
universality\textquotedblright\ assumption, W should treat L with F inside
performing a measurement just as he would treat any other unmeasured quantum
system, in which case the evolution in L is considered to be fully unitary and
before W's measurement the quantum state is%
\begin{equation}
\left\vert \Psi(t)\right\rangle =\left\vert L(t)\right\rangle \left\vert
M(t=0)\right\rangle ,
\end{equation}
where we assumed W's pointer has no evolution previous to the coupling with L.
Alternatively, if W discards the universality assumption and assumes instead
that measurements are always ``special'' (as per von Neumann's rule,
irrespective of where and how the measurement was made), he will take the
Friend's measurement to result in a (possibly effective) collapse of the
global quantum state. In this case W would consider the state of L to be given
not by $\left\vert \Psi(t)\right\rangle $ but by a statistical mixture of
density matrices $\left\vert \mathcal{L}+\right\rangle \left\langle
\mathcal{L}+\right\vert $ and $\left\vert \mathcal{L}-\right\rangle
\left\langle \mathcal{L}-\right\vert $. Wigner in his paper \cite{wigner-r}
concludes that in practice, no experiment could discriminate between both
alternatives.\ Indeed, for a genuine macroscopic size laboratory the
interferences that need to be detected are so small that they become
effectively undetectable \cite{ballentine,matzkin2011}, an argument that goes
back to Bohm \cite{bohm-tb}.

\subsection{EWFS: Two sets of laboratories\label{2L}}

\begin{figure}[tb]
\includegraphics[width=10cm]{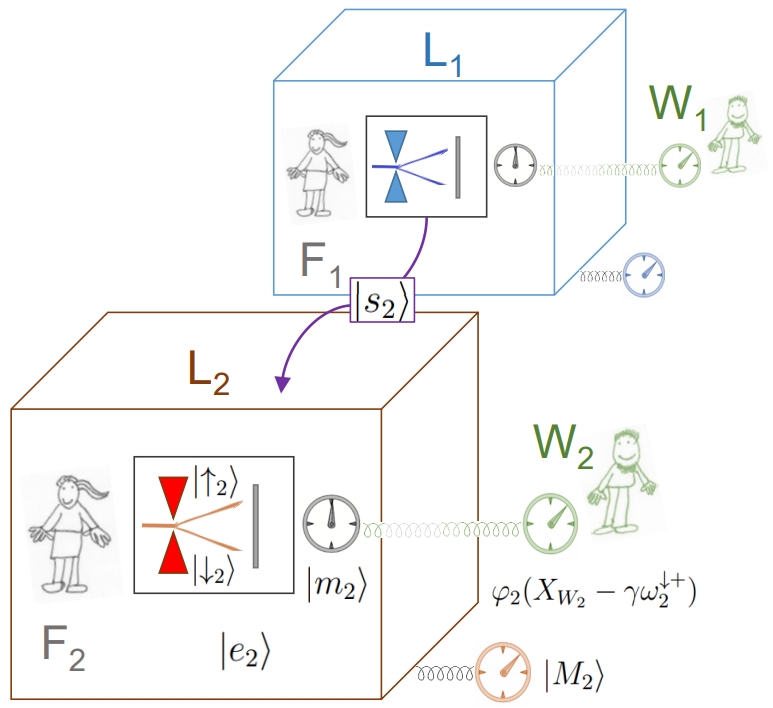}  \caption{Wigner's Extended Friend
Scenario (EWFS). In addition to original WFS of Fig. \ref{fig1}, reproduced
here in the top half of the figure, a Friend F$_{2}$ in a laboratory L$_{2}$
receives a spin in state $\ket{s_2}$, entangled with the spin in L$_{1}$ when
unitary evolution takes place. W$_{2}$ has his own non-invasive probe coupled
to systems inside L$_{2}$.}%
\label{fig2}%
\end{figure}

Drawing on a version of the WFS proposed by Deutsch \cite{deutsch}, Brukner
\cite{bruckner} and Frauchiger and Renner \cite{renner} introduce an Extended
Wigner Friend Setup (EWFS) composed of two sets of Wigner Friend setups
sharing spins 1/2 in an entangled state (see Fig. \ref{fig2}). Our scheme will
follow the latter proposal \cite{renner}, in which a pair of spin 1/2 atoms is
prepared in the non-maximally entangled state%
\begin{equation}
\left\vert \chi\right\rangle =\frac{1}{\sqrt{3}}\left\vert +_{1}\right\rangle
\left\vert \downarrow_{2}\right\rangle +\frac{\sqrt{2}}{\sqrt{3}}\left\vert
-_{1}\right\rangle \left\vert +_{2}\right\rangle \label{es}%
\end{equation}
where $\uparrow\downarrow$ refer to the eigenstates of the spin compoment
along $z,$ $\sigma_{z}\left\vert \uparrow\right\rangle =\left\vert
\uparrow\right\rangle $, $\sigma_{z}\left\vert \downarrow\right\rangle
=-\left\vert \uparrow\right\rangle $ and $\left\vert \pm\right\rangle =\left(
\left\vert \downarrow\right\rangle \pm\left\vert \uparrow\right\rangle
\right)  /\sqrt{2}$ are the eigenstates of $\sigma_{x}$. Employing the
entangled state (\ref{es}) leads to a Bell-type inequality with $0 $ as the
upper bound \cite{hardy,ghirardi}, instead of the usual bound of 2 obtained
with maximally entangled states that Brukner uses in his scheme
\cite{bruckner}. Conceptually, the advantage of having $0$ as an upper bound
is that the violation of the inequality by the quantum version appears as
logically impossible for a classical model in which all the measured
observables have joint values (recall that when observables do not commute,
the quantum formalism precludes the existence of joint probability
distributions \cite{fine}).

In the narrative of Ref. \cite{renner}, F$_{1}$ measures her spin first and
then sends the other atom to L$_{2}$ in state $\left\vert \downarrow
_{2}\right\rangle $ if the outcome was $+$ and in the state $\left\vert
+_{2}\right\rangle $ if F$_{1}$ obtained $-$. This is equivalent to dealing
with Eq. (\ref{es}) from the outset but has the advantage of starting with a
product state for the atoms, which at $t=0$ are both inside $L_{1}$.\ In this
case however L$_{1}$ and L$_{2}$ are not isolated from each other, though our
main requirement is that they should be isolated from the external agents
W$_{1}$ and W$_{2}$.

The initial state of L$_{1}$ is thus%
\begin{equation}
\left\vert L_{1}(t=0\right\rangle =\left\vert s_{1}\right\rangle \left\vert
s_{2}\right\rangle \left\vert m_{1}(t=0)\right\rangle \left\vert
e_{1}(t=0)\right\rangle ,\label{l1t0}%
\end{equation}
where
\begin{equation}
\left\vert s_{1}\right\rangle =\left(  \frac{1}{\sqrt{3}}\left\vert
+_{1}\right\rangle +\frac{\sqrt{2}}{\sqrt{3}}\left\vert -_{1}\right\rangle
\right)  \label{st0}%
\end{equation}
and $\left\vert s_{2}\right\rangle $ can be any spin state. The first step of
F$_{1}$'s measurement involves the unitary coupling (\ref{uc1}) leading to
\begin{equation}
\left\vert L_{1}(0<t<t_{1})\right\rangle =\left(  \frac{1}{\sqrt{3}}\left\vert
+_{1}\right\rangle \left\vert m_{1}+\right\rangle +\frac{\sqrt{2}}{\sqrt{3}%
}\left\vert -_{1}\right\rangle \left\vert m_{1}-\right\rangle \right)
\left\vert s_{2}\right\rangle \left\vert e_{1}(t=0)\right\rangle .
\end{equation}
Upon completion of the measurement at $t=t_{1}$, an observer inside L$_{1}$
reads a definite outcome projecting the pre-measurement state to%
\begin{equation}
\left\vert \mathcal{L}_{1}(t_{1})\right\rangle \left\vert s_{2}\right\rangle
=\left\{
\begin{array}
[c]{c}%
\left\vert \mathcal{L}_{1}+\right\rangle \left\vert \downarrow_{2}%
\right\rangle \equiv\left\vert +_{1}\right\rangle \left\vert m_{1}%
+\right\rangle \left\vert e_{1}+\right\rangle \left\vert \downarrow
_{2}\right\rangle \\
\left\vert \mathcal{L}_{1}-\right\rangle \left\vert +_{2}\right\rangle
\equiv\left\vert -_{1}\right\rangle \left\vert m_{1}-\right\rangle \left\vert
e_{1}-\right\rangle \left\vert +_{2}\right\rangle
\end{array}
\right.  ;\label{proj1}%
\end{equation}
we have assumed a gate coupling the two spins brings $\left\vert
s_{2}\right\rangle $ to either $\left\vert \downarrow_{2}\right\rangle $ or
$\left\vert +_{2}\right\rangle $ depending on whether $\left\vert
+_{1}\right\rangle $ or $\left\vert -_{1}\right\rangle $ were obtained. Spin 2
is sent to L$_{2}$ at $t_{1}^{\prime}$ whose state becomes%
\begin{equation}
\left\vert \mathcal{L}_{2}(t_{1}^{\prime})\right\rangle =\left\{
\begin{array}
[c]{c}%
\left\vert +_{2}\right\rangle \left\vert m_{2}\right\rangle \left\vert
e_{2}+\right\rangle \\
\left\vert \downarrow_{2}\right\rangle \left\vert m_{2}\right\rangle
\left\vert e_{2}\downarrow\right\rangle
\end{array}
\right.  ,\label{L2tp1}%
\end{equation}
where $\left\vert m_{2}\right\rangle $ is F$_{2}^{{}}$'s pointer in the ready
state. F$_{2}$ measures her spin in the $\left\vert \uparrow\downarrow
\right\rangle $ basis, again through the interaction (\ref{uc1}) and after her
measurement is complete at $t=t_{2}$ the state of L$_{2}$ becomes
\begin{equation}
\left\vert \mathcal{L}_{2}(t_{2})\right\rangle =\left\{
\begin{array}
[c]{c}%
\left\vert \mathcal{L}_{2}\uparrow\right\rangle \equiv\left\vert \uparrow
_{2}\right\rangle \left\vert m_{2}\uparrow\right\rangle \left\vert
e_{2}\uparrow\right\rangle \\
\left\vert \mathcal{L}_{2}\downarrow\right\rangle \equiv\left\vert
\downarrow_{2}\right\rangle \left\vert m_{2}\downarrow\right\rangle \left\vert
e_{2}\downarrow\right\rangle
\end{array}
\right.  .\label{proj2}%
\end{equation}
F$_{2}$ has at that point obtained a definite outcome.

Now if we assume that for the external agents W$_{1}$ and W$_{2}$, L$_{1}$ and
L$_{2}$ are ordinary (though macroscopic) quantum systems, the laboratories
evolution is described unitarily and instead of Eq. (\ref{proj1}) we have%
\begin{equation}
\left\vert L_{1}(t_{1})\right\rangle =\frac{1}{\sqrt{3}}\left\vert
+_{1}\right\rangle \left\vert \downarrow_{2}\right\rangle \left\vert
m_{1}+\right\rangle \left\vert e_{1}+\right\rangle +\frac{\sqrt{2}}{\sqrt{3}%
}\left\vert -_{1}\right\rangle \left\vert +_{2}\right\rangle \left\vert
m_{1}-\right\rangle \left\vert e_{1}-\right\rangle .
\end{equation}
Upon receiving atom 2, L$_{2}$ gets entangled with L$_{1}$ and the overall
state at $t_{2}$ is%
\begin{equation}
\left\vert L_{12}(t_{2})\right\rangle =\frac{1}{\sqrt{3}}\left\vert
\mathcal{L}_{1}+\right\rangle \left\vert \mathcal{L}_{2}\downarrow
\right\rangle +\frac{1}{\sqrt{3}}\left\vert \mathcal{L}_{1}-\right\rangle
\left[  \left\vert \mathcal{L}_{2}\uparrow\right\rangle +\left\vert
\mathcal{L}_{2}\downarrow\right\rangle \right]  , \label{l12}%
\end{equation}
where we have used the notation introduced in Eqs. (\ref{proj1}) and
(\ref{proj2}). Finally W$_{1}$ measures L$_{1}$ in the basis $\left\vert
\mathcal{L}_{1}\uparrow\downarrow\right\rangle $ and W$_{2}$ measures L$_{2}$
in the basis $\left\vert \mathcal{L}_{2}\pm\right\rangle $ defined as per the
spin basis by%
\begin{align}
\left\vert \mathcal{L}_{i}\uparrow\downarrow\right\rangle  &  =\frac{1}%
{\sqrt{2}}\left(  \left\vert \mathcal{L}_{i}+\right\rangle \mp\left\vert
\mathcal{L}_{i}-\right\rangle \right) \label{ba1}\\
\left\vert \mathcal{L}_{i}\pm\right\rangle  &  =\frac{1}{\sqrt{2}}\left(
\left\vert \mathcal{L}_{i}\downarrow\right\rangle \pm\left\vert \mathcal{L}%
_{i}\uparrow\right\rangle \right)  . \label{ba2}%
\end{align}
It is useful to rewrite Eq. (\ref{l12}) in the Ws' measurement basis,%
\begin{equation}
\left\vert L_{12}(t_{2})\right\rangle =\frac{1}{\sqrt{12}}\left(  -\left\vert
\mathcal{L}_{1}\uparrow\right\rangle \left\vert \mathcal{L}_{2}+\right\rangle
+\left\vert \mathcal{L}_{1}\uparrow\right\rangle \left\vert \mathcal{L}%
_{2}-\right\rangle +3\left\vert \mathcal{L}_{1}\downarrow\right\rangle
\left\vert \mathcal{L}_{2}+\right\rangle +\left\vert \mathcal{L}_{1}%
\downarrow\right\rangle \left\vert \mathcal{L}_{2}-\right\rangle \right)  .
\label{lu12}%
\end{equation}

The contradiction obtained by Frauchiger and Renner \cite{renner} takes here
the following form.

\begin{enumerate}
\item[(i)] If F$_{2}$ obtains the outcome $\uparrow$, then necessarily F$_{1}
$ has obtained $-$ [by Eq. (\ref{proj1}), with $\left\vert +\right\rangle
=(|\downarrow>+|\uparrow>)/\sqrt{2}$]$;$

\item[(ii)] If W$_{2}$ obtains $-$, then necessarily F$_{1}$ has obtained $+$
[from Eq. (\ref{lu12}) with Eqs. (\ref{ba2}) and (\ref{proj1})]$;$

\item[(iii)] If W$_{1}$ obtains $\uparrow$, then necessarily F$_{2}$ has
obtained $\uparrow$ [from Eq. (\ref{lu12}) with Eqs. (\ref{ba1}) and
(\ref{proj2})];

\item[(iv)] It follows from (i)-(iii) that the outcomes W$_{1}=\uparrow,$
W$_{2}=-$ cannot be obtained jointly\ (indeed, W$_{1}=\uparrow$ implies that
F$_{1}=-,$ but then W$_{2}=-$ cannot be obtained);

\item[(v)] From Eq. (\ref{lu12}), the outcomes W$_{1}=\uparrow,$ W$_{2}=-$ can
be obtained with non-zero probability.
\end{enumerate}

Statements (iv) and (v) are in direct contradiction. Technically, this follows
from the fact that taken together, points (ii)-(iv) implicitly assume the
existence of a joint probability distribution for the outcomes of $\sigma_{x}$
and $\sigma_{z}$ within each set \{L$_{i},$W$_{i}$\}.\ This is well-known to
be impossible, given that $\sigma_{x}$ and $\sigma_{z}$ do not commute: if
only unitary evolution is assumed, the outcome probabilities are obtained by
adding amplitudes that interfere as per (v). But here agents W$_{i}$ still
need to account for the outcomes obtained by the Friends which is the reason
points (ii)-(iv) appear to hold. Note that if global collapse occurs, then
neither (ii) nor (iii) (and therefore (iv)) hold, since the correlations
between the outcomes of W$_{1}$ and F$_{2}$ as those between W$_{2}$ and
F$_{1}$ follow from Eqs. (\ref{proj1}) and (\ref{proj2}).\ We are only adding
probabilities and no contradiction appears.

\section{Monitoring the laboratories with non-invasive minimally disturbing
measurements}

\label{sec3}

We work out here non-invasive minimally disturbing schemes that allow an
external observer to gain information on the quantum states of laboratory
systems or of a laboratory itself. In order to be non-invasive, the readout of
the measurement cannot change the outcomes nor the outcome probabilities of
the Friends or the external agents W. Any type of fuzzy meter \cite{mensky}
with a weak coupling strength could be considered for our scheme.\ For
definiteness we will use weak measurements, a minimally disturbing scheme that
has been widely investigated recently. The underlying conceptual issue --
whether any type of interaction, be it non-invasive, fundamentally changes the
closed character of the laboratories L$_{i}$ will be discussed below in
Sec.\ \ref{discussion}.

\subsection{Weak measurements}

The scheme we use is based on Weak measurements (WM), originally introduced by
Aharonov, Albert and Vaidman \cite{AAV} (see eg \cite{svensson,matzkinFP} for
an introduction to the main concepts). The idea is to extract information
about a given property, represented by an observable $\hat{A}$ on a system
that evolves from a prepared initial state $\left\vert \zeta(t=0)\right\rangle
$ towards the final eigenstate $\left\vert b_{f}\right\rangle $ obtained after
measuring a different observable $\hat{B} $. This is done by creating at some
time $t_{w}$ a von Neumann interaction as per Eq. (\ref{uc1}), but with a very
weak coupling constant (that will be labeled $\gamma$ instead of $g$) so that
the ensuing evolution operator can be expanded to first order as
$1-i\gamma\hat{A}\hat{P}/\hbar$ (as above, $\hat{P}$ is an observable of the
probe to which $\hat{A}$ is coupled to). Then we let the system, now
effectively entangled with the weakly coupled probe, evolve up to $t_{f}$ when
a standard projective measurement of $\hat{B}$ is made.

Up to some time $t_{w}<t<t_{f}$, the system-probe state evolves according to%
\begin{equation}
U(t,t_{w})\left(  1-i\gamma\hat{A}\hat{P}\right)  U(t_{w},0)\left\vert
\zeta(t=0)\right\rangle \left\vert \varphi(t=0)\right\rangle \label{wevol}%
\end{equation}
where $\left\vert \varphi\right\rangle $ denotes the quantum state of the
probe and $U$ is the system evolution operator (for simplicity we assume no
evolution for the probe). Assume that the result of the $\hat{B}$ measurement
at $t_{f}$ is $b_{f}$ and $\left\vert b_{f}\right\rangle $ denotes the
corresponding eigenstate. It is straightforward to show (see, eg
\cite{matzkinFP}) that the weakly coupled probe state after $b_{f}$ has been
obtained is%
\begin{equation}
\left\vert \varphi(t_{f})\right\rangle =\left\langle b_{f}\right\vert \left.
\zeta(t_{f})\right\rangle \exp\left(  -i\gamma A_{f}^{w}\hat{P}\right)
\left\vert \varphi(t=0)\right\rangle \label{fins}%
\end{equation}
where%
\begin{equation}
A_{f}^{w}=\frac{\left\langle b_{f}\right\vert U(t_{f},t_{w})\hat{A}\left\vert
\zeta(t_{w})\right\rangle }{\left\langle b_{f}\right\vert \left.  \zeta
(t_{f})\right\rangle }%
\end{equation}
is known as the weak value of $\hat{A}$. $A_{f}^{w}$ can be complex. If
$\left\vert \varphi(t=0)\right\rangle $ is a Gaussian in position space and
$\hat{P}$ is the momentum Eq. (\ref{fins}) states that the initial probe state
is shifted by $\gamma\operatorname{Re}A_{f}^{w}$. Since $\gamma$ is small the
shift will be very small -- the initial and final pointer states substantially
overlap and in practice the measurements must be repeated in order to gather
enough statistics.

Note that through the weak value, the final probe state depends on the initial
quantum state of the system $\left\vert \zeta\right\rangle $ but also on the
final projective measurement that is made on the system.\ This is a property
that we will see to be useful in the Wigner Friend scenarios. Note also that
the weak interaction modifies the evolution of the system state to first order
in $\gamma$ (compare Eq. (\ref{wevol}) with the free evolution $\left\vert
\zeta(t)\right\rangle =U(t,0)\left\vert \zeta(t=0)\right\rangle $). In
particular the probabilities to obtain an outcome for any subsequent
measurement are not modified to first order (they depend on $\gamma^{2}$). In
this sense the weak measurement of $\hat{A}$ is minimally disturbing.

\subsection{Monitoring the atoms}

\subsubsection{WFS}

Let us consider the original Wigner setup, with a laboratory L$_{1}$, a friend
F$_{1}$\ and an external agent W$_{1}$, and for definiteness assume L$_{1}$ is
initially in state $\left\vert L_{1}(t=0\right\rangle $ as given by Eq.
(\ref{l1t0}), except for the spin $s_{2}$ that plays here no role and will be
left out. We further assume W$_{1}$ measures L$_{1}$ in the $\sigma_{z} $
basis as in Sec.\ \ref{2L} above, with a pointer initially in the ready state
$\left\vert M_{1}(t=0)\right\rangle $. In addition W$_{1}$ has a probe with
momentum variable $\hat{P}_{W_{1}}$, initially in state $\varphi_{1}(X_{W_{1}%
})$ that will be weakly coupled to the spin through an interaction Hamiltonian
$\gamma(t)\hat{\omega}_{1}\hat{P}_{W_{1}}\ $(see Fig. \ref{fig1}), where
$\hat{\omega}_{1}$ is an observable related to the spin (eg, a spin component
along a convieniently chosen axis, or a spin projection operator). The weak
interaction takes place before F measures her spin in the $\sigma_{x}$ basis,
and the initial state evolves for $t<t_{1}$ to
\begin{equation}
\left\vert L_{1}(t)\right\rangle \left\vert \varphi_{1}(t)\right\rangle
=e^{-ig\sigma_{x}\hat{P}/\hbar}e^{-i\gamma\hat{\omega}_{1}\hat{P}_{W_{1}%
}/\hbar}\left\vert s_{1}\right\rangle \left\vert m_{1}(t=0)\right\rangle
\left\vert e_{1}(t=0)\right\rangle \varphi_{1}(X_{W_{1}}) \label{preml1}%
\end{equation}
where $\left\vert s_{1}\right\rangle $ is given by Eq. (\ref{st0}). The
unitary with the weak coupling is expanded to first order and the standard von
Neumann term is expanded in the $\left\vert \pm_{1}\right\rangle $ basis,
yielding%
\begin{equation}
\left[  \frac{1}{\sqrt{3}}\left\vert m_{1}+\right\rangle \left\vert
+_{1}\right\rangle \varphi_{1}(X_{W_{1}}-\gamma\omega_{1}^{+})+\frac{\sqrt{2}%
}{\sqrt{3}}\left\vert m_{1}-\right\rangle \left\vert -_{1}\right\rangle
\varphi_{1}(X_{W_{1}}-\gamma\omega_{1}^{-})\right]  \left\vert e_{1}%
(t=0)\right\rangle . \label{prem1}%
\end{equation}
We have not included the pointer state $\left\vert M_{1}\right\rangle $ that
plays no role at this stage, and we have set $\gamma\equiv\int\gamma
(t^{\prime})dt^{\prime}$ as in the usual von Neumann measurement [see below
Eq. (\ref{VNe})].\ The quantities%
\begin{equation}
\omega_{1}^{\pm}\equiv\frac{\left\langle \pm\right\vert \hat{\omega}%
_{1}\left\vert s_{1}\right\rangle }{\left\langle \pm\right\vert \left.
s_{1}\right\rangle } \label{wvw1}%
\end{equation}
are the weak values. In most instances the exact choice of $\hat{\omega}_{1}$
is not important, as long as it is known to W$_{1}$ so he can infer the weak
value by measuring the probe. Indeed, the dependence of the weak value on the
initial and final states $\left\vert s_{1}\right\rangle $ and $\left\vert
\pm\right\rangle $ is what will matter most in the present context. However
when the choice of $\hat{\omega}_{1}$ becomes significant we will mention it.
Note that some specific choices can be particularly useful in some
circumstances -- for instance $\hat{\omega}_{1}=\Pi_{+}\equiv\left\vert
+\right\rangle \left\langle +\right\vert $ will leave the probe untouched when
the spin outcome is $-$, since by Eq. (\ref{wvw1}) we then have $\left\langle
-\right\vert \Pi_{+}\left\vert s_{1}\right\rangle =0$.

If we assume global collapse upon F's measurement then Eq. (\ref{prem1}) leads
to
\begin{equation}
\left\vert \mathcal{L}_{1}(t_{1})\right\rangle \varphi_{1}(t_{1})=\left\{
\begin{array}
[c]{c}%
\left\vert \mathcal{L}_{1}+\right\rangle \varphi_{1}(X_{W_{1}}-\gamma
\omega_{1}^{+})\\
\left\vert \mathcal{L}_{1}-\right\rangle \varphi_{1}(X_{W_{1}}-\gamma
\omega_{1}^{-})
\end{array}
\right.  \label{coll1}%
\end{equation}
with $\left\vert \mathcal{L}_{1}\pm\right\rangle =\left\vert \pm
_{1}\right\rangle \left\vert m_{1}\pm\right\rangle \left\vert e_{1}%
\pm\right\rangle $. W$_{1}$ finally measures L$_{1}$ in the $\sigma_{z}$ basis
by coupling his pointer in the ready state $\left\vert M_{1}(t=0)\right\rangle
.\ $This is a standard projective measurement that leaves the pointer in state
$\left\vert M_{1}\uparrow\right\rangle $ or $\left\vert M_{1}\downarrow
\right\rangle $. The probe states are shifted by the weak values $\omega
_{1}^{\pm}$ correlated with the states $\left\vert \mathcal{L}_{1}%
\pm\right\rangle .\ $W$_{1}$ therefore obtains one of the following final
states:
\begin{equation}
\left\{
\begin{array}
[c]{c}%
\left\vert M_{1}\uparrow\right\rangle \left\langle \mathcal{L}_{1}%
\uparrow\right\vert \left.  \mathcal{L}_{1}+\right\rangle \varphi_{1}%
(X_{W_{1}}-\gamma\omega_{1}^{+})\\
\left\vert M_{1}\uparrow\right\rangle \left\langle \mathcal{L}_{1}%
\uparrow\right\vert \left.  \mathcal{L}_{1}-\right\rangle \varphi_{1}%
(X_{W_{1}}-\gamma\omega_{1}^{-})
\end{array}
\right.  \qquad\left\{
\begin{array}
[c]{c}%
\left\vert M_{1}\downarrow\right\rangle \left\langle \mathcal{L}_{1}%
\downarrow\right\vert \left.  \mathcal{L}_{1}+\right\rangle \varphi
_{1}(X_{W_{1}}-\gamma\omega_{1}^{+})\\
\left\vert M_{1}\downarrow\right\rangle \left\langle \mathcal{L}_{1}%
\downarrow\right\vert \left.  \mathcal{L}_{1}-\right\rangle \varphi
_{1}(X_{W_{1}}-\gamma\omega_{1}^{-})
\end{array}
\right.  . \label{col1}%
\end{equation}

If instead unitary evolution is assumed, Eq. (\ref{prem1}) becomes%
\begin{equation}
\frac{1}{\sqrt{3}}\left\vert \mathcal{L}_{1}+\right\rangle \varphi
_{1}(X_{W_{1}}-\gamma\omega_{1}^{+})+\frac{\sqrt{2}}{\sqrt{3}}\left\vert
\mathcal{L}_{1}-\right\rangle \varphi_{1}(X_{W_{1}}-\gamma\omega_{1}^{-}).
\label{post1}%
\end{equation}
Note that the laboratory has become slightly entangled with the non-invasive
probe (slightly in the sense that $\gamma$ is small so that the zeroth order
term is the product state $\left(  \left\vert \mathcal{L}_{1}+\right\rangle
+\sqrt{2}\left\vert \mathcal{L}_{1}-\right\rangle \right)  \varphi_{1}\left(
X_{W_{1}}\right)  /\sqrt{3}$). Finally W$_{1}$ measures L$_{1}$ in the
$\sigma_{z}$ basis, so writing Eq. (\ref{post1}) in that basis yields%
\begin{align}
&  \left\vert \mathcal{L}_{1}\uparrow\right\rangle \left[  \frac{1}{\sqrt{6}%
}\varphi_{1}(X_{W_{1}}-\gamma\omega_{1}^{+})-\frac{1}{\sqrt{3}}\varphi
_{1}(X_{W_{1}}-\gamma\omega_{1}^{-})\right] \nonumber\\
&  +\left\vert \mathcal{L}_{1}\downarrow\right\rangle \left\{  \frac{1}%
{\sqrt{6}}\varphi_{1}(X_{W_{1}}-\gamma\omega_{1}^{+})+\frac{1}{\sqrt{3}%
}\varphi_{1}(X_{W_{1}}-\gamma\omega_{1}^{-})\right\}  . \label{unit1}%
\end{align}
Hence when W$_{1}$ obtains $\uparrow$ (or $\downarrow$) the probe will be
found in the states given by the term between $\left[  .\right]  $ and
$\left\{  .\right\}  $ respectively, indicating in both cases a superposition
of the spin in states $\pm$.

The first important result we have obtained is that a weakly coupled probe
allows us to empirically discriminate the mixed pointer states obtained from
Eq. (\ref{col1}) from the interfering patterns of Eq. (\ref{unit1}), whereas
as we have noted at the end of Sec. \ref{1L} this is in practice impossible
for macroscopically interfering states. That said, according to Eq.
(\ref{col1}) the information W$_{1}$ gets from his probe $\varphi_{1}$ if
global collapse\ is assumed matches the outcome that F$_{1}$ announces she
obtained: there is perfect consistency. Instead if pure unitary evolution is
assumed [Eq. (\ref{unit1})] the probe states read by W$_{1}$ indicate a
superposition of the spin in states $+$ and $-$: this is inconsistent with the
fact that F$_{1}$ obtained a definite outcome. Therefore, if we accept the
existence of a non-invasive coupling (this contentious point -- under which
conditions this coupling contradicts the closed character of L$_{1}$, given
that in practical terms it doesn't change the dynamical processes taking place
inside L$_{1}$ -- will be tackled in Sec.\ \ref{discussion}), the original
Wigner Friend scenario leads to a contradiction. The origin of this
contradiction is obvious: W$_{1}$'s probe is non-invasively coupled to the
same spin whose state collapses when F$_{1}$ completes her measurement. Recall
that a collapse on the entangled state (\ref{prem1}) occurs when F$_{1}$
measures the position of her pointer $\left\vert m_{1}\pm\right\rangle $:
depending on the value of $X$, either $m_{1+}(X)$ or $m_{1-}(X)$ will be
non-zero.\ So if F$_{1}$ obtains an outcome, the superposition given by Eq.
(\ref{post1}) cannot occur.

\subsubsection{EWFS}

We now apply the same method to the Extended Wigner Friend scenario of Sec.
\ref{2L}. The external agents W$_{1}$ and W$_{2}$ couple a probe with a weak
interaction to the atoms in laboratories L$_{1}$ and L$_{2}$ respectively (see
Fig. 2). For W$_{1}$, the process is exactly the one we have described
immediately above.$\ $W$_{2}$ also has a probe in state $\varphi_{2}(X_{W_{2}%
})$ that couples to the spin $s_{2}$ of atom 2 once it is received by F$_{2}%
$.\ For definiteness assume the coupling Hamiltonian to be given by
$\gamma(t)\hat{\omega}_{2}\hat{P}_{W_{2}}$ (where the notation follows the
pattern introduced above).

\paragraph{Global collapse upon the Friends measurement}

After F$_{1}$ completes her measurement, a collapse of her meter $m_{1}(X)$
leads to a weakly coupled probe in states $\varphi_{1}(X_{W_{1}}-\gamma
\omega_{1}^{+})$ or $\varphi_{1}(X_{W_{1}}-\gamma\omega_{1}^{-}).$
Concurrently L$_{2}$ is in either of the states given by Eq. (\ref{L2tp1}).
With the weakly coupled probe initially in state $\varphi_{2}(X_{W_{2}})$, the
same analysis \footnote{The term $\gamma\left\vert m\uparrow\right\rangle $
that appears due to the weak coupling is neglected since the probability to
detect this pointer state is second order in $\gamma$.} leading to Eq.
(\ref{coll1}) can be done for L$_{2}$ yielding%
\begin{equation}
\left\vert \mathcal{L}_{2}(t_{2})\right\rangle \left\vert \varphi_{2}%
(t_{2})\right\rangle \left\vert \mathcal{L}_{1}(t_{2})\right\rangle \left\vert
\varphi_{1}(t_{2})\right\rangle =\left\{
\begin{array}
[c]{c}%
\begin{array}
[c]{c}%
\left\vert \mathcal{L}_{2}\uparrow\right\rangle \varphi_{2}(X_{W_{2}}%
-\gamma\omega_{2}^{\uparrow+})\left\vert \mathcal{L}_{1}-\right\rangle
\varphi_{1}(X_{W_{1}}-\gamma\omega_{1}^{-})\\
\left\vert \mathcal{L}_{2}\downarrow\right\rangle \varphi_{2}(X_{W_{2}}%
-\gamma\omega_{2}^{\downarrow+})\left\vert \mathcal{L}_{1}-\right\rangle
\varphi_{1}(X_{W_{1}}-\gamma\omega_{1}^{-})
\end{array}
\\
\left\vert \mathcal{L}_{2}\downarrow\right\rangle \varphi_{2}(X_{W_{2}}%
-\gamma\omega_{2}^{\downarrow\downarrow})\left\vert \mathcal{L}_{1}%
+\right\rangle \varphi_{1}(X_{W_{1}}-\gamma\omega_{1}^{+})
\end{array}
\right.  \label{over12}%
\end{equation}
with $\left\vert \mathcal{L}_{2}\uparrow\right\rangle =\left\vert \uparrow
_{2}\right\rangle \left\vert m_{2}\uparrow\right\rangle \left\vert
e_{2}\uparrow\right\rangle $ and where%
\begin{equation}
\omega_{v}^{s_{f},s_{c}}\equiv\frac{\left\langle s_{f}\right\vert \hat{\omega
}_{2}\left\vert s_{c}\right\rangle }{\left\langle s_{f}\right\vert \left.
s_{c}\right\rangle }.
\end{equation}
$\hat{\omega}_{2}$ is a spin operator or projector related to atom 2,
$\left\vert s_{c}\right\rangle =\left\vert \downarrow\right\rangle $ or
$\left\vert +\right\rangle $ is the spin state of the atom received by F$_{2}$
and $\left\vert s_{f}\right\rangle =\left\vert \downarrow\right\rangle $ or
$\left\vert \uparrow\right\rangle $ is the outcome of F$_{2}$'s measurement.
It can be checked that $\omega_{2}^{\downarrow\downarrow}$ is the expectation
value $\left\langle \downarrow\right\vert \hat{\omega}_{2}\left\vert
\downarrow\right\rangle $. Finally the external observers W$_{i}$ make their
measurements. For W$_{1}$ the outcomes were given in Eq. (\ref{col1}). W$_{2}$
measures L$_{2}$ and the outcomes are $\left\vert \mathcal{L}_{2}%
\pm\right\rangle $ as given by Eq. (\ref{ba2}). His pointer $\left\vert
M_{2}\right\rangle $ is thus correlated with%
\begin{equation}
\left\{
\begin{array}
[c]{c}%
\left\vert M_{2}+\right\rangle \left\langle \mathcal{L}_{2}+\right\vert
\left.  \mathcal{L}_{2}\uparrow\right\rangle \varphi_{2}(X_{W_{2}}%
-\gamma\omega_{2}^{\uparrow+})\\
\left\vert M_{2}+\right\rangle \left\langle \mathcal{L}_{2}+\right\vert
\left.  \mathcal{L}_{2}\downarrow\right\rangle \varphi_{2}(X_{W_{2}}%
-\gamma\omega_{2}^{\downarrow+})\\
\left\vert M_{2}+\right\rangle \left\langle \mathcal{L}_{2}+\right\vert
\left.  \mathcal{L}_{2}\downarrow\right\rangle \varphi_{2}(X_{W_{2}}%
-\gamma\omega_{2}^{\downarrow\downarrow})
\end{array}
\right.  \left\{
\begin{array}
[c]{c}%
\left\vert M_{2}-\right\rangle \left\langle \mathcal{L}_{2}-\right\vert
\left.  \mathcal{L}_{2}\uparrow\right\rangle \varphi_{2}(X_{W_{2}}%
-\gamma\omega_{2}^{\uparrow+})\\
\left\vert M_{2}-\right\rangle \left\langle \mathcal{L}_{2}-\right\vert
\left.  \mathcal{L}_{2}\downarrow\right\rangle \varphi_{2}(X_{W_{2}}%
-\gamma\omega_{2}^{\downarrow+})\\
\left\vert M_{2}-\right\rangle \left\langle \mathcal{L}_{2}-\right\vert
\left.  \mathcal{L}_{2}\downarrow\right\rangle \varphi_{2}(X_{W_{2}}%
-\gamma\omega_{2}^{\downarrow\downarrow})
\end{array}
\right.  . \label{col2}%
\end{equation}

The joint outcomes for W$_{1}$ and W$_{2}$ are independent and can be obtained
from Eqs. (\ref{col1}), (\ref{over12}) and (\ref{col2}).\ They can read from
their probes $\varphi_{i}$ the outcomes obtained by their respective friends.
For example let us look at the case W$_{1}=\uparrow,$ W$_{2}=-$ which as we
saw lies at the basis of the contradiction [see points (iv)-(v) below Eq.
(\ref{lu12})]. W$_{1}=\uparrow$ can be obtained with either F$_{1}=+$ or $-$
[left handside of Eq. (\ref{col1})].\ W$_{2}=-$ can be obtained with either
F$_{2}=\uparrow$ or $\downarrow$ with F$_{1}$ having obtained either $+$ or
$-$.\ Hence none of the points (ii)-(iv) hold in this case.\ For instance if
we take $\hat{\omega}_{2}=\left\vert -\right\rangle \left\langle -\right\vert
$, then $\omega_{2}^{\uparrow+}=\omega_{2}^{\downarrow+}=0$ and W$_{2}$ will
see a shift in the weak probe only when the state received by F$_{2}$ was
$\left\vert \downarrow\right\rangle $ (and hence F$_{1}$ obtained $+$), while
for $\hat{\omega}_{2}=\left\vert \uparrow\right\rangle \left\langle
\uparrow\right\vert $, $\omega_{2}^{\downarrow+}=\omega_{2}^{\downarrow
\downarrow}=0$ only when F$_{2}$ received the atom in state $\left\vert
+\right\rangle $ and obtained the outcome $\uparrow$.

\paragraph{Unitary evolution of the laboratories\label{fue}}

If L$_{1}$ and L$_{2}$ are asumed to evolve unitarily (as per the
\textquotedblleft universality\textquotedblright\ assumption) the weak
couplings with W$_{1}$ and W$_{2}$'s non-invasive probes are taken into
account by the evolution operator $e^{-i\gamma\hat{\omega}_{2}\hat{P}_{W_{2}%
}/\hbar}e^{-i\gamma\hat{\omega}_{1}\hat{P}_{W_{1}}/\hbar}$ so that the overall
wavefunction given by Eq. (\ref{l12}) becomes with this coupling
\begin{align}
\left\vert \Psi(t_{2})\right\rangle  &  =\frac{1}{\sqrt{3}}\left\vert
\mathcal{L}_{1}+\right\rangle \left\vert \mathcal{L}_{2}\downarrow
\right\rangle \varphi_{1}(X_{W_{1}}-\gamma\omega_{1}^{+})\varphi_{2}(X_{W_{2}%
}-\gamma\omega_{2}^{\downarrow\downarrow})+\nonumber\\
&  +\frac{1}{\sqrt{3}}\left\vert \mathcal{L}_{1}-\right\rangle \varphi
_{1}(X_{W_{1}}-\gamma\omega_{1}^{-})\left[  \left\vert \mathcal{L}_{2}%
\uparrow\right\rangle \varphi_{2}(X_{W_{2}}-\gamma\omega_{2}^{\uparrow
+})+\left\vert \mathcal{L}_{2}\downarrow\right\rangle \varphi_{2}(X_{W_{2}%
}-\gamma\omega_{2}^{\downarrow+})\right]  . \label{uniwv}%
\end{align}
Let \ us look at the correlation between outcomes as by points (i)-(v) below
Eq. (\ref{lu12}):

(i$^{\prime}$) Point (i) [F$_{2}=\uparrow\Rightarrow$F$_{1}=-$] is verified by
construction so it holds for all the cases we have examined so far.

(ii$^{\prime}$) If we write Eq. (\ref{uniwv}) in the basis $\left\vert
\mathcal{L}_{1}\pm\right\rangle \left\vert \mathcal{L}_{2}\pm\right\rangle $,
the term proportional to $\left\vert \mathcal{L}_{2}-\right\rangle $ between
the square brackets $[..]$ is $-\varphi_{2}(X_{W_{2}}-\gamma\omega
_{2}^{\uparrow+})+\varphi_{2}(X_{W_{2}}-\gamma\omega_{2}^{\downarrow+})$ for
which the leading term vanishes. We thus neglect it and we are left with the
sole term $\left\vert \mathcal{L}_{1}+\right\rangle \left\vert \mathcal{L}%
_{2}-\right\rangle \varphi_{1}(X_{W_{1}}-\gamma\omega_{1}^{+})\varphi
_{2}(X_{W_{2}}-\gamma\omega_{2}^{\downarrow\downarrow})$\ coming from the
upper line. This implies that (ii) holds [W$_{2}=-\Rightarrow$F$_{1}=+$] and
the weak probe $\varphi_{2}$ further indicates to W$_{2}$ that F$_{2}$
received and measured the atom spin in state $\downarrow$.

(iii$^{\prime}$) We write Eq. (\ref{uniwv}) in the basis $\left\vert
\mathcal{L}_{1}\updownarrow\right\rangle \left\vert \mathcal{L}_{2}%
\updownarrow\right\rangle $; to leading order, we are left with $\left\vert
\mathcal{L}_{1}\uparrow\right\rangle \left\vert \mathcal{L}_{2}\uparrow
\right\rangle \varphi_{1}(X_{W_{1}}-\gamma\omega_{1}^{-})\varphi_{2}(X_{W_{2}%
}-\gamma\omega_{2}^{\uparrow+})$. Hence (iii) holds [W$_{1}=\uparrow
\Rightarrow$F$_{2}=\uparrow$] and we have the information through the weak
probes that the outcome for F$_{1}$ was $-$.

(iv$^{\prime}$) The information carried by the weak probes in (ii') and (iii')
strengthens the apparent contradiction (iv) pointed out above: (ii$^{\prime}$)
tells us that W$_{2}=-\Rightarrow($F$_{1}=+$ and F$_{2}=\downarrow)$ while
from (iii') W$_{1}=\uparrow\Rightarrow($F$_{1} =-$ and F$_{2} =\uparrow$).

(v$^{\prime}$) Eq. (\ref{uniwv}) is written in the basis $\left\vert
\mathcal{L}_{1}\uparrow\downarrow\right\rangle \left\vert \mathcal{L}_{2}%
\pm\right\rangle $.\ The term yielding the outcomes W$_{1}=\uparrow,$
W$_{2}=-$ is factored by%
\begin{align}
&  \left\vert \mathcal{L}_{1}\uparrow\right\rangle \left\vert \mathcal{L}%
_{2}-\right\rangle \left\{  \varphi_{1}(X_{W_{1}}-\gamma\omega_{1}^{+}%
)\varphi_{2}(X_{W_{2}}-\gamma\omega_{2}^{\downarrow\downarrow})\right.
\nonumber\\
&  \left.  +\varphi_{1}(X_{W_{1}}-\gamma\omega_{1}^{-})\varphi_{2}(X_{W_{2}%
}-\gamma\omega_{2}^{\uparrow+})-\varphi_{1}(X_{W_{1}}-\gamma\omega_{1}%
^{-})\varphi_{2}(X_{W_{2}}-\gamma\omega_{2}^{\downarrow+})\right\}  .
\label{vp}%
\end{align}
Here the terms that can be neglected to leading order in $\gamma$ depend on
the choice of $\hat{\omega}_{1}$ and $\hat{\omega}_{2}$.\ In typical cases,
$\omega_{2}^{\downarrow\downarrow}\neq\omega_{2}^{\downarrow+}$ and the second
line becomes negligible to leading order in $\gamma$ leaving the term
$\varphi_{1}(X_{W_{1}}-\gamma\omega_{1}^{+})\varphi_{2}(X_{W_{2}}-\gamma
\omega_{2}^{\downarrow\downarrow})$ corresponding to F$_{1}=+$ and
F$_{2}=\downarrow$. However if $\omega_{2}^{\downarrow\downarrow}=\omega
_{2}^{\downarrow+}$ (eg, for $\hat{\omega}_{2}=\sigma_{z}$) Eq. (\ref{vp})
becomes to leading order%
\begin{equation}
\left\vert \mathcal{L}_{1}\uparrow\right\rangle \left\vert \mathcal{L}%
_{2}-\right\rangle \frac{1}{2}\left(  \varphi_{1}(X_{W_{1}}-\gamma\omega
_{1}^{+})\varphi_{2}(X_{W_{2}}-\gamma\omega_{2}^{\downarrow\downarrow
})+\varphi_{1}(X_{W_{1}}-\gamma\omega_{1}^{-})\varphi_{2}(X_{W_{2}}%
-\gamma\omega_{2}^{\uparrow+})\right)  . \label{vp2}%
\end{equation}
In this case, when W$_{1}=\uparrow,$ W$_{2}=-$ is obtained, the probes are,
like in Eq. (\ref{unit1}) for the original setup, in a state of superposition
of states in which $\left(  \text{F}_{1}=+,\text{F}_{2}=\downarrow\right)  $
and $\left(  \text{F}_{1}=-,\text{F}_{2}=\uparrow\right)  $.\ The probes do
not indicate to the external observers W$_{i}$ that a specific outcome has
been obtained.

We will see below that if the probes are coupled to the pointer or the
environment inside each L$_{i}$ (hence after each Friend's measurement has
occurred), the behavior represented by Eq. (\ref{vp2}) is the generic one. We
therefore defer its discussion to the paragraph immediately below. We wish to
emphasize here that we are combining projective measurements and weak
measurements.\ The projective measurements carried out in points (ii$^{\prime
}$), (iii$^{\prime}$) and (v$^{\prime}$) are not compatible (since they
involve non-commuting observables). The weak measurements are non-invasive
unitary interactions that give information on an observable. Therefore the
problem of compatibility with the final projective measurement does not arise
-- this is actually one of the reasons weak measurements have found to be
useful -- but it should be stressed that weak values are conditioned on how
the amplitudes interfere upon a final projective measurement.

\subsection{Monitoring inside the laboratory}

We investigate the case in which the external agents W$_{i}$ weakly couple an
external probe to an element inside L$_{i}$ rather than to the atoms after
F$_{i}^{\prime}s$ measurement has occurred.\ By \textquotedblleft
element\textquotedblright\ we mean a system that carries information about the
measurement, such as the pointers, or the environment inside L$_{i}$. In the
version in which the Friend's measurement induces an overall collapse a weak
probe of this type would be redundant, as it would simply confirm the outcome
obtained by each of the Friends' projective measurements. So we will only
consider weak coupling in the unitary evolution scheme that would follow from
Wigner's \textquotedblleft universality\textquotedblright\ assumption. We will
simplify the notation by using $\updownarrow$ to stand for \textquotedblleft%
$\uparrow$ or $\downarrow$\textquotedblright.

\subsubsection{WFS}

For definiteness, assume the weakly coupled probe, placed outside L$_{1}$
interacts, through a minute hole, with the measurement apparatus in the
otherwise sealed laboratory. This interaction takes place after the spin
interacted with the Friend's pointer.\ Hence, with $\left\vert s_{1}%
\right\rangle $ yet again given by Eq. (\ref{st0}), the quantum state after
the weak interaction is similar to\ Eq. (\ref{preml1}),%
\begin{equation}
\left\vert L_{1}(t)\right\rangle \left\vert \varphi_{1}(t)\right\rangle
=e^{-i\gamma\hat{\omega}_{1}\hat{P}_{W_{1}}/\hbar}e^{-ig\sigma_{x}\hat
{P}/\hbar}\left\vert s_{1}\right\rangle m_{1}(X,t=0)\left\vert e_{1}%
(t=0)\right\rangle \varphi_{1}(X_{W_{1}})\label{inside}%
\end{equation}
but $\hat{\omega}_{1}$ is now an obervable coupled to the measurement
apparatus after the spin measurement has taken place and the appartus states
have become $\left\vert m_{1}\pm\right\rangle $. W$_{1}$ measures L$_{1}$ in
the \ $\left\vert \mathcal{L}_{1}\updownarrow\right\rangle $ basis, leading
to
\begin{equation}
\ \left[  \frac{1}{\sqrt{3}}\left\langle \mathcal{L}_{1}\updownarrow
\right\vert \left.  \mathcal{L}_{1}+\right\rangle \varphi_{1}(X_{W_{1}}%
-\gamma\omega_{1}^{\updownarrow+})+\frac{\sqrt{2}}{\sqrt{3}}\left\langle
\mathcal{L}_{1}\updownarrow\right\vert \left.  \mathcal{L}_{1}-\right\rangle
\varphi_{1}(X_{W_{1}}-\gamma\omega_{1}^{\updownarrow-})\right]  \left\vert
M_{1}\updownarrow\right\rangle .\label{monicol1}%
\end{equation}
Since $\left\langle \mathcal{L}_{1}\updownarrow\right\vert \left.
\mathcal{L}_{1}\pm\right\rangle =\pm1/\sqrt{2}$, we have essentially the same
result as when the probe was weakly coupled to the spin, see Eq.
(\ref{unit1}). For both outcomes $\updownarrow$, the prediction according to
unitary evolution is that W$_{1}$ will find his weakly coupled probe
$\varphi_{1}$ as indicating a superposition of the spin in states $\pm$. So
W$_{1}$ can again conclude there is an inconsistency between the information
obtained by measuring  $\varphi_{1}$ and the definite outcome F$_{1}$
announced she measured.\ Hence W$_{1}$ can discriminate empirically this
prediction obtained by assuming unitary evolution from the mixed state
expected if the Friend's spin measurement induces a global collapse.

Note that in principle the weakly interacting probe can be coupled to the
environment inside L$_{1}$, or more precisely to the fraction of the
environment affected by the Friend's measurement (assuming we can factor the
environment into an interacting part, described by $\left\vert e_{1}%
\right\rangle $ and a non-interacting part that plays no role), for example
the photons that carry the measurement outcome information. Eq. (\ref{inside})
still holds, with $\hat{\omega}_{1}$ now coupled to the environment that can
quickly be described by states $\left\vert e_{1}\pm\right\rangle $ correlated
with the measurement apparatus states $\left\vert m_{1}\pm\right\rangle $ (we
are neglecting here the dynamics during the very short time it takes for the
environment states to become orthogonal in the pointer state basis). While
remaining orthogonal, the states $\left\vert e_{1}\pm\right\rangle $ change
rapidly with time (see Ref. \cite{relano} for a model of the environment in
the WFS scenario), so that the weak values $\omega_{1}^{\updownarrow+}$ will
depend on the timing of the weak interaction. Hence while Eq. (\ref{monicol1})
still holds if unitary evolution is assumed and the conclusions for W$_{1}$
remain identical to the ones that hold for a probe coupled to the measurement
apparatus, it is difficult to make general statements on the practical
realizability of the weak probe reading without a specific model of the environment.

\subsubsection{EWFS}

Now both W$_{1}$ and W$_{2}$ couple a probe to the measuring apparatus inside
L$_{1}$ and L$_{2}.\ $The non-invasive probes interact with the respective
measurement apparatus after the spin measurements have taken place, the
apparatus being  respectively in states $\left\vert m_{1}\pm\right\rangle $
and $\left\vert m_{2}\pm\right\rangle $.\ Assuming unitary evolution, the
quantum state reads%
\begin{equation}
\left\vert \Psi(t_{2})\right\rangle =e^{-i\gamma\hat{\omega}_{1}\hat{P}%
_{W_{1}}/\hbar}e^{-i\gamma\hat{\omega}_{2}\hat{P}_{W_{2}}/\hbar}\left\vert
L_{12}(t_{2})\right\rangle \varphi_{1}(X_{W_{1}})\varphi_{2}(X_{W_{2}})
\end{equation}
where $\left\vert L_{12}\right\rangle $ is given by Eqs. (\ref{l12}) or
(\ref{lu12}). The same projective measurements as in Sec. \ref{fue} are
performed by W$_{1}$ and W$_{2}$ on $\left\vert \Psi(t_{2})\right\rangle $.
The weak values are now defined relative to the pointer states $\left\vert
m_{1}\right\rangle $ or $\left\vert m_{2}\right\rangle $:%
\begin{equation}
\omega_{i}^{\beta\alpha}=\frac{\left\langle m_{i}\beta\right\vert \hat{\omega
}_{i}\left\vert m_{i}\alpha\right\rangle }{\left\langle m_{i}\beta\right\vert
\left.  m_{i}\alpha\right\rangle },
\end{equation}
where $\alpha$ labels the state in the measurement basis of the Friend F$_{i}$
and $\beta$ a state in the basis in which W$_{i}$ measures. Points
(i$^{\prime\prime}$)-(iv$^{\prime\prime}$) are the same as in the case in
which the non-invasive probe was coupled directly to the spins. Point
(v$^{\prime\prime}$) is slightly different:

(i$^{\prime\prime}$) Again, this point [F$_{2}=\uparrow\Rightarrow$F$_{1}=-$]
is verified by construction.

(ii$^{\prime\prime}$) If we take the scalar products $\left\langle
\mathcal{L}_{1}\pm\right\vert \left\langle \mathcal{L}_{2}-\right\vert $ to
leading order only the term $\left\langle \mathcal{L}_{1}+\right\vert
\left\langle \mathcal{L}_{2}-\right\vert \left.  \Psi\right\rangle $ is
non-zero with the probes left in states $\varphi_{1}(X_{W}-\gamma\omega
_{1}^{++})\varphi_{2}(X_{W_{2}}-\gamma\omega_{2}^{-\downarrow}),$ implying
[W$_{2}=-\Rightarrow$F$_{1}=+$ and F$_{2}=\downarrow$], exactly as in (ii').

(iii$^{\prime\prime}$) Computing the scalar products $\left\langle
\mathcal{L}_{1}\uparrow\right\vert \left\langle \mathcal{L}_{2}\updownarrow
\right\vert \left.  \Psi\right\rangle $ leaves the leading term with the
probes in state $\varphi_{1}(X_{W}-\gamma\omega_{1}^{\uparrow-})\varphi
_{2}(X_{W_{2}}-\gamma\omega_{2}^{\uparrow\uparrow})$. Hence (iii') holds
[W$_{1}=\uparrow\Rightarrow$F$_{2}=\uparrow$ and F$_{1}=-$].

(iv$^{\prime\prime}$) We have the same contradiction as in (iv').

(v$^{\prime\prime}$) The projection $\left\langle \mathcal{L}_{1}%
\uparrow\right\vert \left\langle \mathcal{L}_{2}-\right\vert $ leads\ to
\begin{equation}
\varphi_{1}(X_{W}-\gamma\omega_{1}^{\uparrow+})\varphi_{2}(X_{W_{2}}%
-\gamma\omega_{2}^{-\downarrow})+\varphi_{1}(X_{W}-\gamma\omega_{1}%
^{\uparrow-})\varphi_{2}(X_{W_{2}}-\gamma\omega_{2}^{-\uparrow})-\varphi
_{1}(X_{W}-\gamma\omega_{1}^{\uparrow-})\varphi_{2}(X_{W_{2}}-\gamma\omega
_{2}^{-\downarrow}).
\end{equation}
To lowest order in $\gamma$, this can be put under the form%
\begin{equation}
\frac{1}{2}\left(  \varphi_{1}(X_{W}-\gamma\omega_{1}^{\uparrow+})\varphi
_{2}(X_{W_{2}}-\gamma\omega_{2}^{-\downarrow})+\varphi_{1}(X_{W}-\gamma
\omega_{1}^{\uparrow-})\varphi_{2}(X_{W_{2}}-\gamma\omega_{2}^{-\uparrow
})\right)  .
\end{equation}
This is similar to Eq. (\ref{vp2}) of point (v'), but now holds for any choice
of $\hat{\omega}_{1}$ and $\hat{\omega}_{2}$ concerning the measurement
pointers: the outcomes W$_{1}=\uparrow,$ W$_{2}=-$ can be obtained with
non-zero probability, and when this happens the probes $\varphi_{i}$ indicate
that the Friends' pointers where in a state of superposition of outcomes
$\left(  \text{F}_{1}=+,\text{F}_{2}=\downarrow\right)  $ and $\left(
\text{F}_{1}=-,\text{F}_{2}=\uparrow\right)  $. The probes act here as a
witness of the entanglement of $\left\vert L_{12}(t_{2})\right\rangle $ [Eq.
(\ref{lu12})]. Since W$_{1}$ and W$_{2}$ can communicate, a clever choice of
$\hat{\omega}_{1}$ and $\hat{\omega}_{2}$ would allow them to determine that
their probes have been entangled by the unitary evolution.\ For the agents
W$_{i}$ such a finding would be inconsistent given that the Friends announced
they have obtained definite outcomes (in which case their states should not be entangled).

\section{Discussion\label{discussion}}

\subsection{Summary}

Let us start by summarizing the results we have obtained.

In the original WFS, if the external agent W assumes L evolves unitarily, he
will run into a contradiction: a non-invasive probe coupled (and therefore
entangled with) the Friend's atom or her pointer would display a signature of
the macroscopic superposition. Indeed W can measure the non-invasive probe
irrespective of whether F tells him she obtained a definite outcome. If
unitary evolution is assumed, quantum theory predicts that the probe will
indicate a superposition of the atom spin or pointer states: this is
incompatible with F having obtained a definite outcome. Instead if global
collapse is assumed, quantum theory predicts that the weakly coupled probes
will indicate the outcome obtained by F. Note that while measuring the state
of an entire macroscopic laboratory in order to discriminate a mixed state
from unitary evolution is experimentally out of reach, measuring non invasive
weakly coupled probes could be achieved with present day technologies.

In the EWFS, assuming unitary evolution results in entangled probes.\ This is
a signature that the atoms or the pointers to which the probes were coupled
are entangled. But this is impossible since F$_{1}$ and F$_{2}$ have announced
they have obtained definite outcomes. This inconsistency when unitary
evolution is assumed is similar to the one for the original WFS, with the
superposition in the extended version taking the specific form of
entanglement. This inconsistency is similar to the logical contradiction
obtained by Frauchiger and Renner \cite{renner} [see points (i)-(v) below Eq.
(\ref{lu12})], except that here this contradiction appears as information the
agents could read on their probes. The agents W$_{i}$ would therefore conclude
by reading their probes that they were wrong in assuming unitary evolution, as
the probes are consistent with the Friends' announcements only if global
collapse is assumed.

\subsection{Closed laboratories and non-invasive measurements}

According to the summary we have just given, an external agent should apply
the projection postulate (\textquotedblleft global collapse\textquotedblright)
and not assume unitary evolution for a closed laboratory in which a
measurement takes place, as this will lead him in to an observable
contradiction.\ One may question however whether the laboratories can still be
qualified as closed when a very weak interaction is allowed with the external world.

From a purely logical viewpoint, this objection makes sense: an interaction,
however small, breaks the purported independence of the closed laboratory from
the external world.\ The very idea of attempting to discriminate
observer-dependent from observer independent facts \cite{bruckner} seems
indeed ruined. This is a crucial point in Deutsch's approach \cite{deutsch}
that employs a notion called \textquotedblleft kinematic
independence\textquotedblright\ to assume two subsystems can evolve as
independent worlds.

From the physical viewpoint we adopt here however, this logical independence
is a fiction: a laboratory can never be perfectly isolated. Operationally, a
laboratory is said to be sealed if the inevitable interactions with the rest
of the Universe can be neglected relative to the effects that are
investigated. Here, the sealed character of the laboratory implies that the
operations carried out by the Friends while running their experiment inside
the lab cannot be detected by any type of measurement in the external
Universe, except through the channel opened for the non-invasive probe. The
non-invasive interaction itself is asymptotically small and does not change in
any essential way the dynamical evolution inside the laboratories: in
particular the predictions made inside the laboratories and the outcome
probabilities are not modified. From this perspective, adding non-invasive
probes does not fundamentally alter the physical interactions and processes
inside the laboratories. Our operational assumption is that such a probe
simply records the local value of the system observable it is interacting
with, with minimal change to the system's state. This is how the effect of
weak measurements are generally interpreted \cite{matzkinFP}. 

Another way to tackle this question, that we examined very recently
\cite{AD1}, is to associate a measurement outcome with the existence of a
stable material record. For practical or fundamental reasons, on which we will
not speculate here, a stable material record is immune to linear
superpositions (in Feynman's words, amplitudes are lost and play no more role
\cite{chapel}). We have argued \cite{AD1} that the material record includes
F's memory (or any printout that would record the measurement), so that if the
stable records produced by F's measurement could be erased (through unitary
evolution) by W's subsequent measurement, there is no way to ensure F's
measurement has even existed. This is precisely the situation measured by the weakly coupled probes when unitary evolution is assumed: there is a lack of records that measurements should produce. Hence within this perspective, there
can be no observer-dependent facts even if the sealed laboratory is assumed to evolve unitarily, unless one supposes that the presence of a non-invasive interaction prevents the formation of a stable
material record  for the Friends that would take place for totally sealed laboratories (admittedly, this sounds
quite unlikely given the properties of weak measurements).

\subsection{Measurements: Quantum Mechanics is consistent in its
inconsistency}

Wigner Friend scenarios are at the crossroad of two distinct but related
problems: what constitutes a measurement, and whether quantum mechanics holds
in the macroscopic limit. Any account of the WFS and the problems that it
might raise will hinge on one's stance on these issues, which to a large
extent depends on the interpretation one has in mind (this includes a set
assumptions some of which might be clearly expressed explicitly, along with
others that might be fuzzily implicitly presupposed).

Wigner, shortly after he published the WFS\ argument, endorsed more clearly
the idea that quantum mechanics applied to macroscopic bodies but not to
conscious observers, for whom unitary evolution \textquotedblleft is not
credible\textquotedblright\ (Sec.\ 12 of \cite{wigner69}). Obviously, this
position conflicts with the idea that quantum mechanics should apply in a
\textquotedblleft universal\textquotedblright\ manner to macroscopic systems,
the assumption that is at the root of the WFS and more particularly in the
EWFS of Refs \cite{bruckner,renner}: the agents W$_{i}$ would discard that the
Friends evolved unitarily once they completed their measurements (and
therefore no contradiction will be obtained between the agents).

Purely unitary approaches (such as the Many-Worlds Interpretation) uphold that
the projection postulate makes quantum theory non-universal and logically
inconsistent \cite{deutsch}. In such approaches, whether the Friends and the
agents W evolve or not in different worlds, observing or not different facts
and how different facts can be reconciled when all the agents meet depends on
subtle assumptions concerning the branching mechanism (for a recent discussion
detailing on how branching issues in EPR or GHZ setups might lead agents
evolving independently to reconcile their viewpoints, see \cite{cai}).

In most other approaches, the projection postulate for all the agents must be
implicitly or explicitly employed. It is then straightforward to see
\cite{laloe} that, while the application of the projection postulate remains
ambiguous, all the agents involved in the EWFS can agree beforehand on how to
apply it and avoid reaching contradictions. It has also been argued
\cite{zukowski} that the sole unitary coupling is insufficient to define a
measurement -- unitary evolution leads at best to the premeasurement step of
the measurement process, but this does not suffice in order to obtain recorded
outcomes. Then all the agents will agree on their measurements. Note that the
collapse can be implicit, like when taking the partial trace over the
environment in a decoherence based approach (a detailed decoherence based
account of the WFS and EWFS has recently been given \cite{relano}), or in the
projectors defining consistent histories (such an approach for the EWFS has
also been given \cite{lombardi}).

For approaches in which the state vector is considered to embody an objective
physical reality, global collapse upon measurement for all the agents is the
only option. This is obvious in dynamical collapse theories \cite{grw} in
which a non-linear term is added to to the Schrödinger equation, preventing
unitary evolution. While we do not know of any detailed rendering of the WFS
with dynamical collapse, once the measurement is completed our results
assuming global collapse in Sec. \ref{sec3} will be recovered. In the de
Broglie-Bohm approach, the dual particle-wave ontology renders the effective
collapse somewhat less obvious, as the particle is detected but the
wavefunction -- assumed to be unique for all the agents -- presents (in most
versions of this approach) empty branches that still evolve, creating an
ambiguity over when they remain dynamically relevant
\cite{holland,matzkin-bil}. While slightly different, the recent de
Broglie-Bohm accounts of the WFS and EWFS \cite{sudbery,dustin,drezet} all
agree that particle detection by the Friends changes the Bohmian particle
configurations and trajectories for the external agents W, so that no
contradiction is obtained.

The conclusion we can draw is that Quantum Mechanics is consistent in its
inconsistency: even if we have good reasons to believe that quantum mechanics
should apply to macroscopic bodies, a special rule, involving a global
effective collapse relying on an ambiguous \textquotedblleft shifty
split\textquotedblright\ \cite{bellAM} remains necessary. Quantum measurements
are special and the theoretical accounts remain unsatisfactory, if not
logically inconsistent \cite{deutsch,bellAM}, but operationally there is no
inconsistency in accounting for the observed outcomes.

\section{Conclusion}

\label{sec5}

In this paper, we have investigated Wigner Friend scenarios in which the
external agents make non-invasive measurements inside the otherwise sealed
laboratories. In the original WFS, an external agent W who would want to test
the Friend's unitary evolution would not be able to infer anything by directly
measuring the quantum states of the macroscopic laboratory. With a
non-invasive probe weakly coupled to the Friend's atom or measuring apparatus
-- with a coupling so weak that it does not affect the outcomes nor their
probabilities inside the laboratory -- W will be able to infer the state of
the laboratory after the Friend's measurement. Within the limitations of our
scheme -- in which the Friend's laboratory cannot said to be perflectly closed
-- unitary evolution can be empirically tested, and probably invalidated,
since if this hypothesis is assumed, we are lead to a contradiction between
the Friend having obtained a definite outcome and the state of the probe.

For the extended version of the WFS recently introduced by Brukner
\cite{bruckner} and Frauchiger and Renner \cite{renner}, we have shown that
the non-invasive probes given to each external agent would be entangled if the
hypothesis held.\ But we have seen that if the Friends obtained definite
outcomes, the probes cannot be entangled (this forms the basis of the
Bell-type inequality that leads to the contradiction put forward in Ref
\cite{renner}). Again, in our framework with non-invasive probes, this becomes
a question that can be given an experimental confirmation.

To conclude, we have investigated Wigner Friend scenarios with non-invasive
probes. While there is room to support the idea that the measurement problem
makes quantum theory logically inconsistent, we can rely on the non-invasive
probes to avoid inconsistency at the operational level. This approach could
moreover be confirmed by experiments.

\end{document}